\documentclass[a4paper]{jpconf}

\usepackage[T2A]{fontenc}
\usepackage{amsmath,amssymb,amsfonts,amsthm}
\usepackage{euscript}
\usepackage{graphicx}
\usepackage{wrapfig}

\begin{document}

\title{Bound orbits near black holes with scalar hair}

\author{I M Potashov, Ju V Tchemarina, and A N Tsirulev}

\address{Faculty of Mathematics, Tver State University, 35 Sadovyi, Tver, Russia, 170002}

\ead{potashov.im@tversu.ru, chemarina.yv@tversu.ru, tsirulev.an@tversu.ru}

\begin{abstract}
   We consider spherically symmetric black holes with minimally coupled scalar fields and concentrate our attention on asymptotically flat self-gravitating configurations having the event horizons located at radii much smaller than $2m$. We think of such configurations as rigorous mathematical models of the gravitating objects, surrounded by dark matter, in the centres of normal galaxies. It turns out that the radius of the event horizon of a scalar field black hole always less than the Schwarzschild radius of vacuum black hole of the same mass and can be arbitrary close to zero. In astronomical observations, a key role in distinguishing between black holes, wormholes, and naked singularities plays measuring parameters of bound quasiperiodic orbits, in particular, the shape of an orbit and the angle of precession of its pericentre. We consider a typical two-parameter family of compact scalar field black holes and compute numerically the shapes of some bound orbits. We find that a key feature of bound orbits around a compact black hole is that the angle between closest pericentre points is either negative or, at least, less than that for the Schwarzschild black hole of the same mass.
\end{abstract}

\section{Introduction}

One of the main problems of modern astrophysics is the identification of self-gravitating objects in the centers of galaxies~\cite{Akiyama2015, Fish2016, Goddi2017}. Namely, are they black holes, wormholes, naked singularities, or maybe boson stars? In any of these cases, the observations of bound orbits in the centers of galaxies play a key role for this question~\cite{Li2014, Vieira2014, DeLaurentis2018, Stashko2018, Mishra2018, Willenborg2018, Kratovitch2017}). In this paper, we study some properties of bound orbits near black holes in the model of a static, self-gravitating, spherically symmetric scalar field. The field is supposed to be minimally coupled to gravity and the spacetime to be asymptotically flat. This model is motivated by the modern astronomical observations of the centers of galaxies. Strongly gravitating objects in galactic centers cannot be regarded as being vacuum because they are surrounded by dark matter. In our approach, we model dark matter by a self-interacting scalar field. It is very probably that scalar fields exist in nature, but even in the opposite case, we have a powerful degree of freedom to construct a phenomenological model of dark matter. It is true, at least, for the spherically symmetric objects, because the self-interaction potential can be chosen arbitrarily in the wide class of physically admissible (particularly, positive at infinity~\cite{Nikonov2008}) potentials. In the other words, we can approximately model any spherically symmetric distribution of matter in a self-consistent manner.

In fact, the form of the potential and the distribution of dark matter near the centre should be found from the astronomical observations. Therefore, we use the so-called inverse problem method~\cite{BechmannLechtenfeld1995, BronnikovShikin2002, Tchemarina2009, Azreg-Ainou2010, Solovyev2012} for static, spherically symmetric, scalar field configurations. This method allows us to deal with, in some sense, all the physically admissible potentials at the same time. Our aim is to study some distinctive features of the geodesic motion of massive test particles near the horizons of scalar field black holes. We restrict our attention to compact black holes, which have the radius of the event horizon much smaller than the corresponding Schwarzschild radius and, consequently, show most distinctive features of the shape of bound orbits. In Section~\ref{Sec2} we describe the mathematical background of the inverse problem method. Section~\ref{Sec3} is devoted to an analytical formulation of a simple two-parameter family of compact black holes. In Section~\ref{Sec4} we compare the shapes of bounded orbits around a typical compact black hole with those around the Schwarzschild black hole of the same mass.

In this paper we adopt the metric signature $(+\,-\,-\,-)$ and use the geometrical system of units with $G=1,\,c=1$.

\section{Spherically symmetric scalar field black holes}\label{Sec2}

Self-gravitating minimally coupled scalar fields are described by the action
\begin{equation}\label{action}
\Sigma=\frac{1}{8\pi}\int\! \left(-\frac{1}{2}S - \langle d\phi,d\phi\rangle-2V(\phi)\right) \sqrt[]{|g|}\,d^{\,4}x\,,
\end{equation}
where $S$ is the scalar curvature, $\phi$ the scalar field, $V(\phi)$ the self-interaction potential, and the angle brackets denote the scalar product with respect to the metric. We will write the metric of a static spherically symmetric spacetime in the Schwarzschild-like coordinates (that is, $r$ is the area coordinate) in the form
\begin{equation}\label{metric}
    ds^2=A dt^2- \frac{\,dr^2}{f}- r^{2}(d\theta^2+\sin^{2}\!\theta\, d\varphi^2),
\end{equation}
where the metric functions $A$ and $f$, as well as the field $\phi$, depend only on the radial coordinate. Any static spherically symmetric solution of the Einstein-Klein-Gordon equations for the action~(\ref{action}) with an arbitrary physically admissible potential $V(\phi)$ obeys the quadrature formulae~\cite{Solovyev2012}
\begin{equation}\label{A-f}
A(r)=2r^{2}\!\!\int\limits_{\!r}^{\,\,\infty} \frac{\,\xi-3m}{\,r^4}\,\mathrm{e}^{F}dr\,, \quad f(r)=\mathrm{e}^{-2F}A\,,
\end{equation}
\begin{equation}\label{F-xi}
F(r)=-\!\int\limits_{\!r}^{\,\,\infty}\! {\phi'}^{2}rdr\,,\quad \xi(r)=r+\int\limits_{\!r}^{\,\,\infty}\!\! \left(1-\mathrm{e}^{F}\right)\!dr\,,
\end{equation}
\begin{equation}\label{V}
\widetilde{V}(r)=\frac{1}{2r^2}\!\left(1-3f+ r^2{\phi'}^{2}\!f+ 2\,\mathrm{e}^{-F}\,\frac{\,\xi-3m}{r}\right),
\end{equation}
where a prime denotes differentiation with respect to $r$, $\widetilde{V}(r)=V(\phi(r))$, and the positive parameter $m$ is the Schwarzschild mass. The 'inverse problem method' consists of specifying a monotonic function $\phi(r)$ and finding successively the functions $\mathrm{e}^{F(r)}$, $\xi(r)$, $A(r)$, $f(r)$, $\widetilde{V}(r)$, and the potential $V(\phi)=\widetilde{V}(r(\phi))$. The quadratures~(\ref{A-f})\,--\,(\ref{V}) give us, in some sense, a general solution of the Einstein-Klein-Gordon equations.

We assume that $\phi(r)$ is of class $C^2\!\left[(0,\infty)\right]$ and has the asymptotic behaviour
\begin{equation}\label{phi} \phi=O\!\left(r^{-1/2-\textstyle{\alpha}}\right)\!,\;\;\, r\rightarrow\infty\;\;\;(\alpha>0).
\end{equation}
The functions $F(r)$, $\mathrm{e}^{F(r)}$, and $\xi(r)$ obey the obvious relations
\begin{equation}\label{cond1}
F'>0,\;\;\; \xi>r,\;\;\; 0<\,\xi'=\mathrm{e}^{F}\leqslant\,1,\;\;\; \xi''\!=\left(\mathrm{e}^{F}\right)'\geqslant0\;\;\; \mbox{for all}\;\;r>0,
\end{equation}
\begin{equation}\label{cond2}
\mathrm{e}^{F}\!=1+o\big(r^{-1}\big),\quad \xi\!=r+o(1),\;\;\; r\rightarrow\infty,
\end{equation}
\begin{equation}\label{cond3}
\mathrm{e}^{F}= \mathrm{e}^{F(0)}+o(r),\;\;\; \xi\!=\xi(0)+\xi'(0)r+o(r),\;\; r\rightarrow0,
\end{equation}
where
\begin{equation}\label{cond4}
 \mathrm{e}^{F(0)}\equiv \lim\limits_{r\rightarrow0}\mathrm{e}^{F(r)},\; \quad\xi(0)>0,\quad\;\; \xi'(0)=\mathrm{e}^{F(0)}\geqslant0.
\end{equation}
Note also that $\mathrm{e}^{F(0)}>0$ if and only if $\phi(r)$ grows slower than $\ln{r}$ as $r\rightarrow0$.
Substituting the expansions (\ref{cond2}) and (\ref{cond3}) into the quadrature (\ref{A-f}) and using the conditions (\ref{cond1}) and (\ref{cond4}), we find that $A(r)$ has the asymptotic behaviours
\begin{equation}\label{A-asympt}
A(r)=1-\frac{3m}{r}\,+\,o(1/r),\;\; r\rightarrow\infty, \qquad
A(r)=\frac{2}{3}\:\! \frac{\xi(0)-3m}{r}\,\mathrm{e}^{F(0)}\,+\, O(1),\;\; r\rightarrow0.
\end{equation}
It means, first, that $m$ is the Schwarzschild mass and, second, that \textit{this metric function defines a black hole if and only if}\footnote{It defines a naked singularity if $3m<\xi(0)$, while if $3m=\xi(0)$, we obtain either a regular solution or a naked singularity.}
\begin{equation}\label{xi(0)<3m}
3m>\xi(0).
\end{equation}
Furthermore, it follows directly from the quadrature (\ref{A-f}) and the second inequality in (\ref{cond1}) that for a given scalar field \textit{the event horizon radius} $r_{\!\scriptscriptstyle h}$ \textit{of a black hole with mass} $m>\xi(0)/3$ \textit{is less then the corresponding Schwarzschild radius}, that is, $r_{\!\scriptscriptstyle{}h}<2m$.

In what follows, we will consider \textit{compact black holes} for which $3m\rightarrow\xi(0)+0$ and, consequently, $r_{\!\scriptscriptstyle{}h}\ll{}2m$ (see Fig~\ref{Orbits1}).

\section{A family of compact black holes}
\label{Sec3}

In this section we consider a typical two-parameter family of compact black holes; the assertion about typicality is based on many numerical and analytical experiments with other suitable examples of spherically symmetric scalar fields. In the purely analytical treatment of the quadratures~(\ref{A-f})\,--\,(\ref{V}), it is more convenient to start with some specially chosen function~$\mathrm{e}^{F(r)}$, because the functions $\mathrm{e}^{F(r)}$, $\xi(r)$, and $\phi$ uniquely determine each other.

The strictly monotonic piecewise analytic function
\begin{equation}\label{expF} \mathrm{e}^{F(r)}=1-\alpha+\frac{\,2\alpha}{7}\!\;r^3,\;\; 0\leqslant{}r\leqslant{}1, \quad\mbox{and}\quad \mathrm{e}^{F(r)}=1-\frac{\,2\alpha}{r^3}+ \frac{\,9\alpha}{7r^4},\;\; 1\leqslant{}r<\infty
\end{equation}
has continuous derivatives up to second order at the point $r=1$. By direct integration in (\ref{F-xi}) and then in (\ref{A-f}), we obtain
\begin{equation}\label{xi}
\xi= \frac{\,3\alpha}{2}+(1-\alpha)\!\;r+ \frac{\,\alpha\,}{14}\!\;r^4,\;\; 0\leqslant{}r\leqslant{}1, \quad\mbox{and}\quad \xi= r+\frac{\,\alpha\,}{r^2}-\frac{\,3\alpha}{7r^3}\,,\;\; 1\leqslant{}r<\infty,
\end{equation}
\begin{multline}
A=-\frac{(1-\alpha)(2m-\alpha)}{r}+(1\!-\!\alpha)^2\,+ \\
\left(\frac{6}{7}\!\;(2m-\alpha)\ln{}r+ \frac{8}{5}-\frac{88}{105}\!\; \alpha-\frac{54}{49}m\!\right)\! \alpha{}r^2-\frac{5}{7}\!\;\alpha(1-\alpha)r^3- \frac{\alpha^2}{98}\!\;r^6,\;\;\;\;\; 0\leqslant{}r\leqslant{}1, \qquad\;\;\nonumber
\end{multline}
\begin{equation}\label{A}
\quad A= 1\,-\,\frac{2m}{r}\,-\,\frac{2\alpha}{5r^3}\,+\, \left(\!m+\frac{1}{7}\right)\!\frac{2\alpha}{r^4}\,-\, \frac{54\alpha{}m}{49r^5}\,-\, \frac{\alpha^2}{2r^6}\,+\, \frac{10\alpha^2}{21r^7}\,-\, \frac{27\alpha^2}{245r^8}\,, \;\; 1\!\leqslant{}\!r\!<\!\infty, \;\;
\end{equation}
where $\alpha\!\in\![0,1)$ is the parameter of 'intensity' of the scalar field. The field can be obtained by numerical solving of the problem $\phi'=\sqrt{F'/r\,},\,\,\phi(\infty)=0$; at infinity $\phi\sim \sqrt{8\alpha/3}\,r^{-3/2}$. Note that the length scale in (\ref{expF})\,--\,(\ref{A}) is fixed by choosing the matching point at $r=1$. In studying bound orbits, we are interested only in the metric functions $A$ and $f$ and can, therefore, use an arbitrary unit of length. On the other hand, the solution~(\ref{A-f})\,--\,(\ref{V}) is invariant under the scale transformations
\begin{equation}
r\rightarrow r/a,\quad m\rightarrow m/a,\quad V\rightarrow a^2V,\quad\;\;a>0.\nonumber
\end{equation}
It means that by applying $a=m$ in this transformation, one can choose, as it is usually done in general relativity,  the mass of a black hole as the current unit of length.

\begin{figure}[!h]
  \;\;\begin{minipage}{0.47\textwidth}
    \includegraphics[width=0.97\textwidth, height=6.81cm]{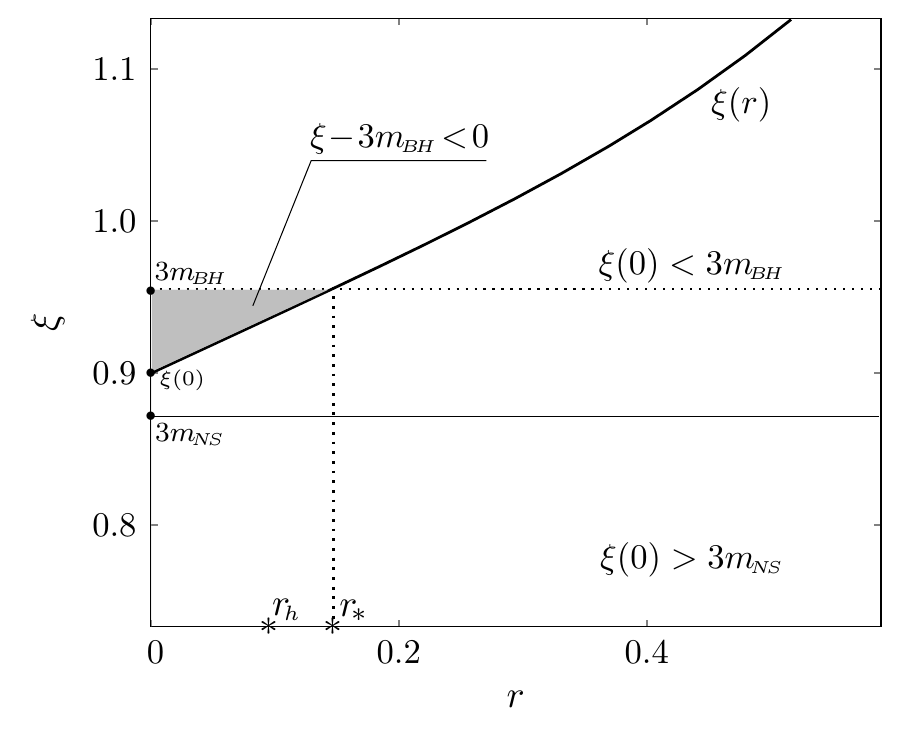}
  \end{minipage}\qquad
  %%%%%%%%%%%%%%%%%%%%%%%
 \begin{minipage}{0.47\textwidth}
 \vspace{-2ex}
    \includegraphics[width=0.98\textwidth, height=6.91cm]{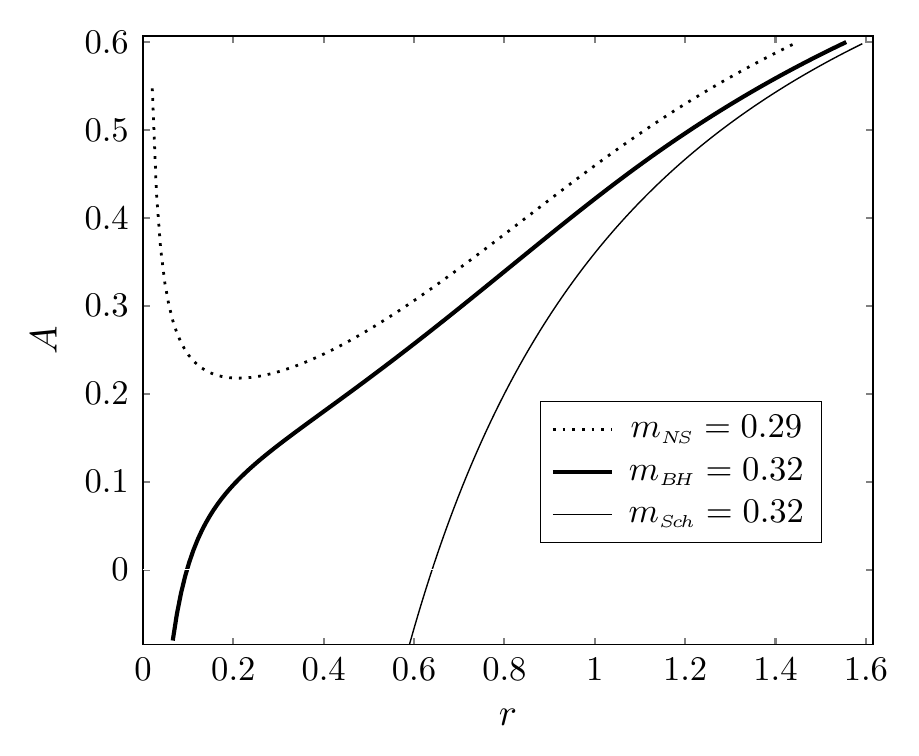}
  \end{minipage}
  \caption{ The functions $\xi(r)$ (left) and $A(r)$ (right). Parameters:
  $r_{\!*}=0.15$, $r_{\!\scriptscriptstyle h}=0.01$,  $\alpha=0.6$, $m_{\!\scriptscriptstyle BH}=0.32$, $m_{\!\scriptscriptstyle Sch}=0.32$,
  $m_{\!\scriptscriptstyle NS}=0.29$.}
  \label{Orbits1}
\end{figure}

We see that the radius of the horizon $r_{\!\scriptscriptstyle h}\!\rightarrow\!0$\; as\; $3m\rightarrow\xi(0)\!+\!0$ ($\xi(0)=3\alpha/2$) or, equivalently, as $m\rightarrow\alpha/2\!+\!0$. For compact black holes, one can estimate directly from~(\ref{A}) that $r_{\!\scriptscriptstyle h}\approx(2m-\alpha)/(1-\alpha)$. \textit{In what follows, we choose} $\alpha=0.6$. For $m=m_{\!\scriptscriptstyle B\!H}=\alpha/2+0.02=0.32$, we obtain a compact black hole with $r_{\!\scriptscriptstyle h}\approx0.01$. The corresponding metric function $A_{\scriptscriptstyle B\!H}$, as well as the Schwarzschild solution $A_{\scriptscriptstyle Sch}$ of the same mass and the naked singularity solution $A_{\scriptscriptstyle N\!S}$ with $\alpha=0.6$ and $m=0.29$, are shown in Fig.~\ref{Orbits1}. In the region $r>3r_{\!\scriptscriptstyle Sch}=1.92$ (outside the Schwarzshcild innermost stable circular orbits), where we also have $\mathrm{e}^{F(r)}\!\!\:\big|_{\alpha=0.6}\approx1$, all the three spacetime geometries practically coincide.

\section{Bound orbits near the event horizons of compact black holes}
\label{Sec4}

\begin{figure}[!h]
  %\centering
  \begin{minipage}{0.49\textwidth}
    \includegraphics[width=0.97\textwidth, height=7.31cm]{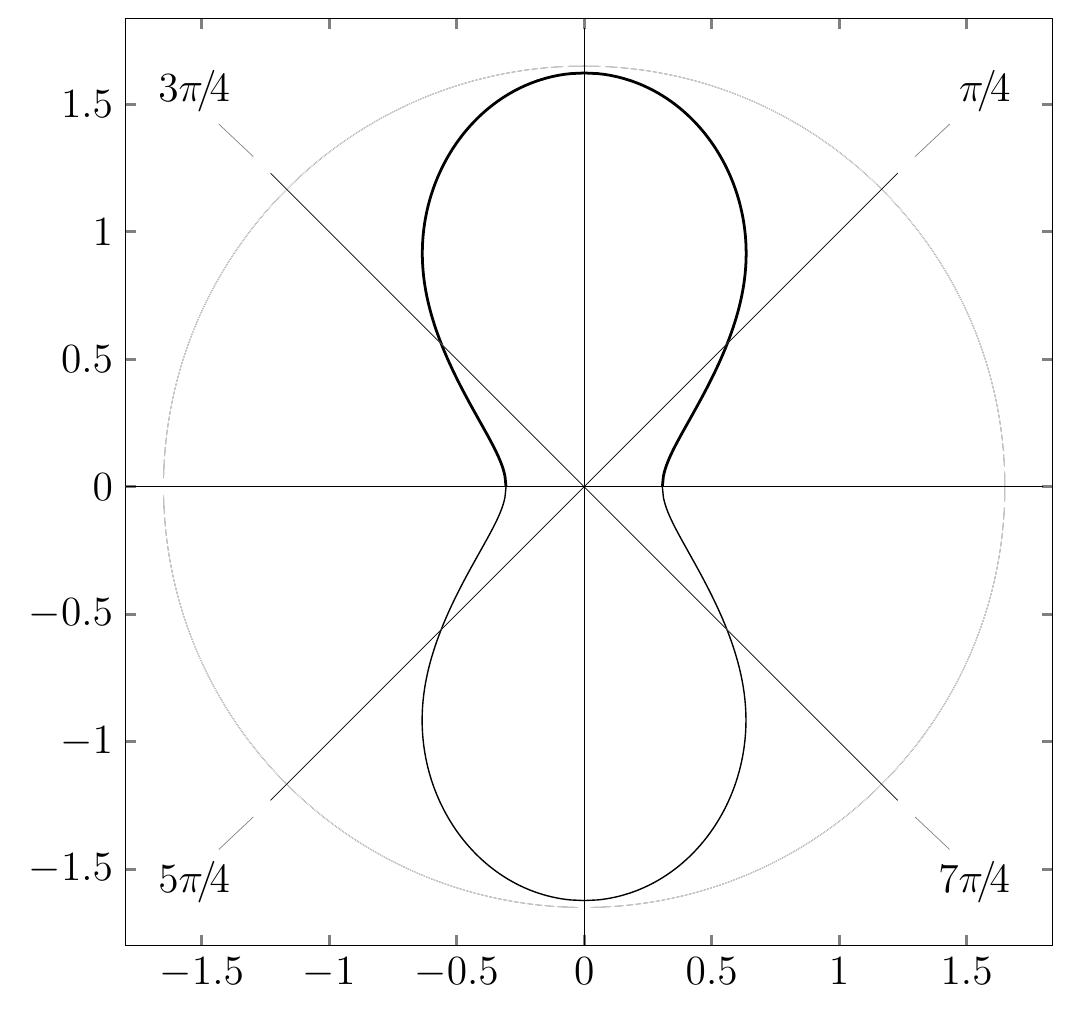}\\
  \end{minipage}
  %%%%%%%%%%%%%%%%%%%%%%%
 \begin{minipage}{0.49\textwidth}
  \vspace{-3ex}
    \includegraphics[width=0.97\textwidth, height=7.31cm]{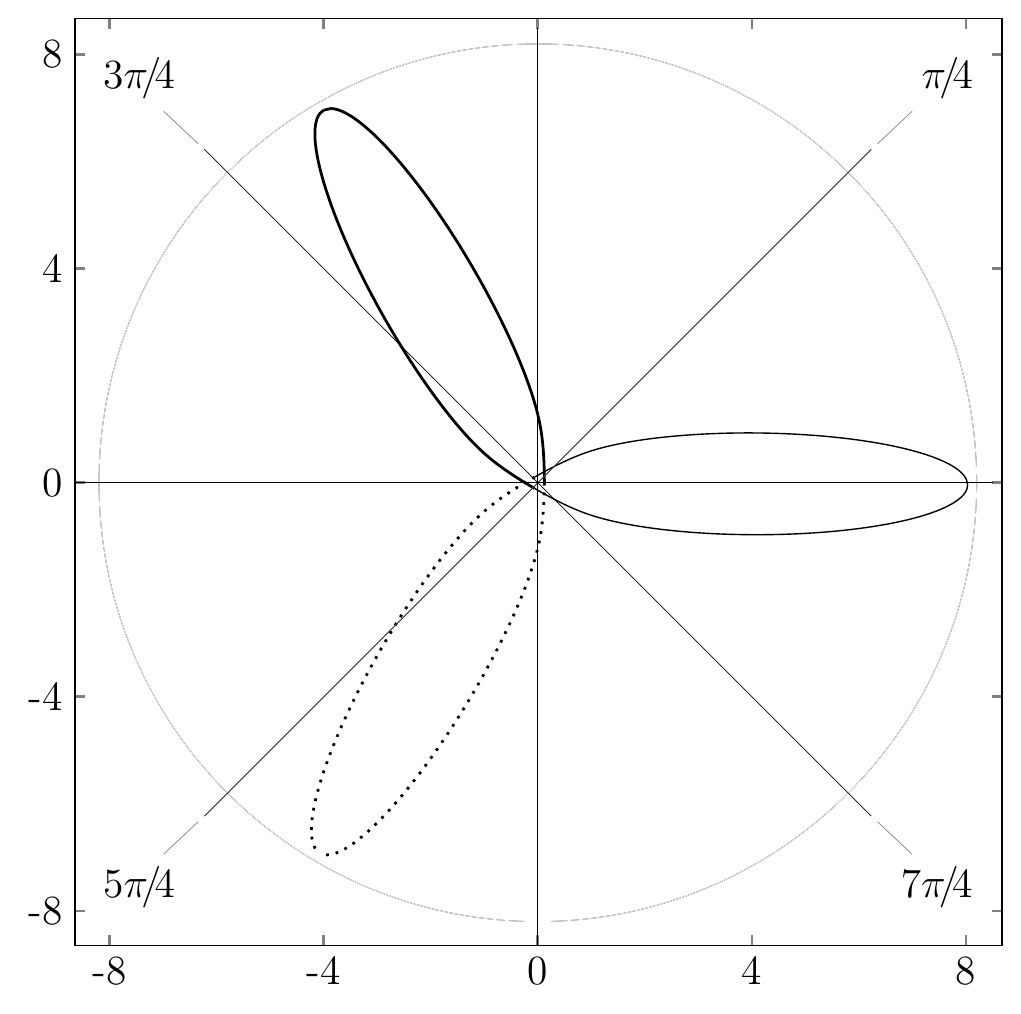}
  \end{minipage}
  %#######################################################
  %#######################################################
  \begin{minipage}{0.49\textwidth}
    \includegraphics[width=0.97\textwidth, height=7.41cm]{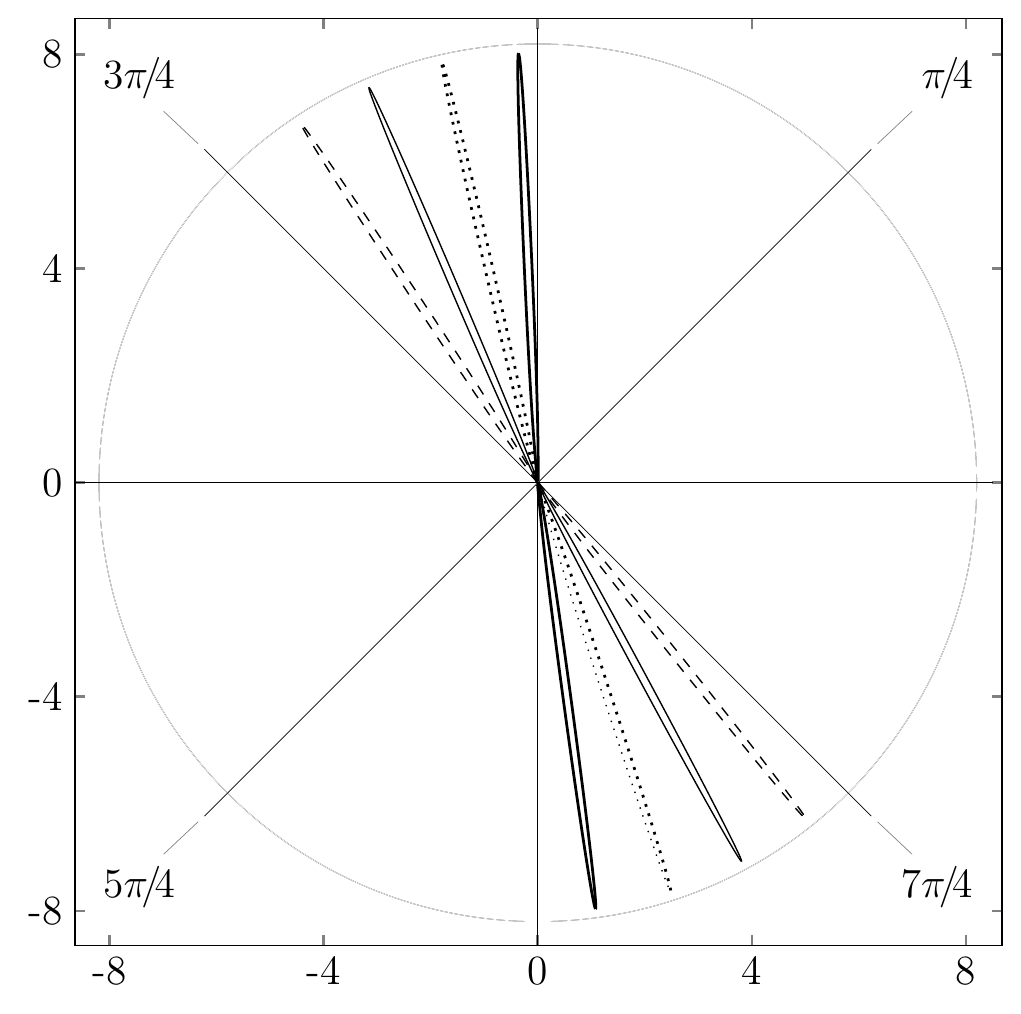}
  \end{minipage}\;\,
  %%%%%%%%%%%%%%%%%%%%%%%
 \begin{minipage}{0.49\textwidth}
    \includegraphics[width=0.97\textwidth, height=7.41cm]{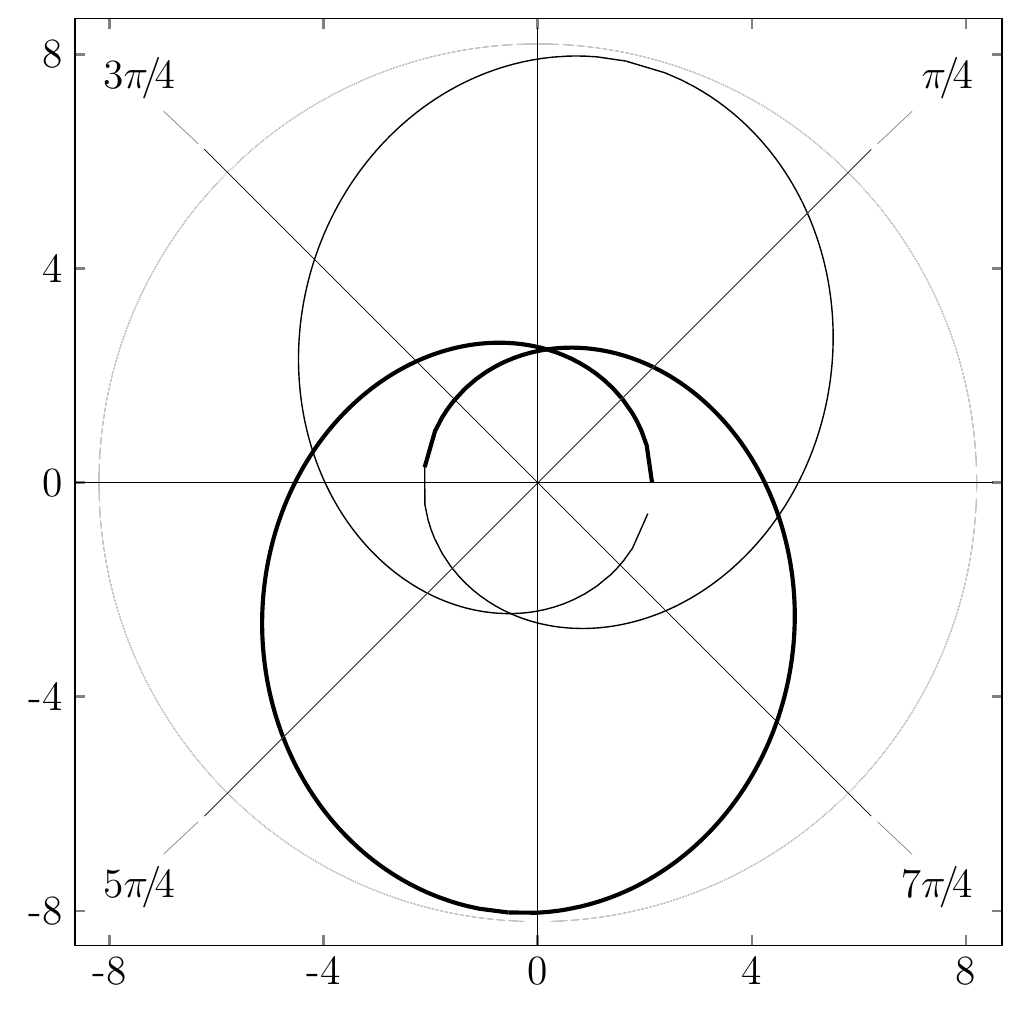}
  \end{minipage}
  \caption{Top-left panel: $\alpha=0.6$, $m=\alpha/2+0.02=0.32$, $\Delta\varphi=-\pi$, $r_{\!\scriptscriptstyle min}=0.31$, $r_{\!\scriptscriptstyle max}=1.62$, $J=0.6$, $E^2=0.689$. In the other plots, $\alpha=0.6$, $r_{\!\scriptscriptstyle max}\approx8.04$.
  Top-right panel: $m=\alpha/2+10^{-4}=0.3001$, $r_{\!\scriptscriptstyle h}=5\cdot10^{-4}$, $\Delta\varphi=-2\pi/3$, $J=0.262$, $E^2=0.926$. Bottom-left panel: $m=\alpha/2+10^{-5}=0.30001$, $r_{\!\scriptscriptstyle h}=5\cdot10^{-5}$, $\Delta\varphi=-3.05$, $J=0.02$, $E^2=0.925$.   Bottom-right panel: the Schwarzschild black hole, $m=0.3001$, $\Delta\varphi=+3.01$, $J=1.2$, $E^2=0.946$.}
  \label{Orbits2}
\end{figure}

In a spherically symmetric spacetime, the Lagrangian for geodesics is constant along each geodesic and does not depend explicitly on the coordinates $t$ and $\varphi$. We write the corresponding integrals of motion as
\begin{equation}\label{int}
\frac{dt}{ds}=\frac{E}{A}\,,\qquad
\frac{d\varphi}{ds}=\frac{J}{\,r^2}\,,\qquad
\left(\frac{dr}{ds}\right)^{\!2}=  \mathrm{e}^{-2F}\! \left(E^2- {V\vphantom{\underline{A}}}_{\!\!eff}\right), \quad
{V\vphantom{\underline{A}}}_{\!\!eff}= A\!\left(1+\frac{\,J^2}{\,r^2}\right),
\end{equation}
where $E$ is the specific energy and $J$ the specific angular momentum of a massive free particle. The effective potential of a black hole spacetime vanishes at the horizon, tends to unity as $r\rightarrow\infty$, and has, for sufficiently large $J$, at least one minimum and one maximum outside the horizon. We are primarily interested in the shape of a bound orbit and the angle of precession $\Delta\varphi$ of the orbit, which can be expressed by the obvious relations
\begin{equation}\label{precess}
\varphi_{\!{}_{\scriptscriptstyle osc}}=2J\!\int \limits_{\,r_{\!\scriptscriptstyle min}}^{\:r_{\!\scriptscriptstyle max}}\! \frac{\,\mathrm{e}^{F}} {\,r^2\sqrt{E^2-V_{\!\scriptscriptstyle eff}\,}}\,dr\,,\quad \Delta\varphi\,=\,\varphi_{\!{}_{\scriptscriptstyle osc}}-2\pi\,,
\end{equation}
where $r_{\!\scriptscriptstyle min}$ and $r_{\!\scriptscriptstyle max}$ are solutions of the equation $E^2-V\vphantom{\underline{A}}_{\!\!eff}=0$ which are uniquely determined by the requirement of being located most nearest to (and on the opposite sides of) the minimum of the effective potential. Thus, a bound orbit of the general type oscillates near a stable circular orbit and $\varphi_{\!{}_{\scriptscriptstyle osc}}$ is the angle between two successive oscillations.

There is \textit{a key difference between bound orbits around a compact black holes and the Schwarzschild black hole of the same mass: in the former case, the angle between closest pericentre points is either negative or, at least, (much) less than that in the latter case.} Note that a bound orbit around a vacuum black hole always has the advance of its pericentre. Fig.~\ref{Orbits2} represents typical trajectories of a test particle. An orbit with the negative angle of precession $\Delta\varphi=-\pi$ is shown in the top-left panel. Two prolate bound orbits (with one and the same apocentre radius) around an extremely compact black hole are represented in the top-right and bottom-left panels. A bound orbit (with approximately the same apocentre radius) around the Schwarzschild black hole of the same mass is shown in the bottom-right panel. It should be stressed that these shapes of bound orbits and their precession angle deficits are typical for compact black holes.

\section{Conclusions}

We consider compact spherically symmetric black holes with minimally coupled scalar fields and concentrate our attention on asymptotically flat self-gravitating configurations having the event horizons located at radii much smaller than $2m$. Such configurations can be thought of as rigorous mathematical models of the strongly gravitating objects, surrounded by dark matter, in the centres of normal galaxies. We show that the radius of the event horizon of a scalar field black hole always less (possibly much less) than the Schwarzschild radius of vacuum black hole of the same mass and can be arbitrary close to zero. We also consider bound orbits around compact black holes and find that the angle between closest pericentre points of such orbits is either negative or, at least, less than that for the Schwarzschild black hole of the same mass. The numerical experiments show that the corresponding radius of the innermost stable circular orbit is also less than in the vacuum case. These features play key roles in distinguishing between scalar field black holes, vacuum black holes, wormholes, naked singularities, and boson stars. Note that boson stars have no innermost stable circular orbits, and scalar field wormholes and naked singularities (of the same positive mass) have marginal (or degenerated) stable circular orbits with the zero angular momentum of test particles. These result can possibly be applied to interpretations of future astronomical observations. In particular, one can hope that direct observation of the central region of $Sgr A^{*}$, will soon be obtained by the Event Horizon Telescope Collaboration.


\section*{References }
\begin{thebibliography}{99}

\bibitem{Akiyama2015}
Akiyama K et al 2015 \textit{Astrophys. J.} \textbf{820}, 90 (\textit{arXiv:} 1602.05527)

\bibitem{Fish2016}
Fish V L et al 2016 \textit{Astrophys. J.} \textbf{820}, 150 (\textit{arXiv:} 1505.03545)

\bibitem{Goddi2017}
Goddi G et al 2017 \textit{Int. J. Mod. Phys.} \textbf{D20}, 1730001-239 (\textit{arXiv:} 1606.088879)

\bibitem{Li2014}
Li Z and Bambi C 2014 \textit{Phys. Rev. D} \textbf{90}, 024071 (\textit{arXiv:} 1405.1883)

\bibitem{Vieira2014}
Vieira R S S, Schee J, Klu\'{z}niak W, Stuchl\'{\i}k Z, and Abramowicz M 2014 \textit{Phys. Rev. D} \textbf{90}, 024035

\bibitem{DeLaurentis2018}
De Laurentis M, Younsi Z, Porth O, Mizuno Y, and Rezzolla L 2018 \textit{Phys. Rev. D} \textbf{97}, 104024

\bibitem{Stashko2018}
Stashko O S and Zhdanov V I 2018 \textit{Gen. Relat. Gravit.} \textbf{50}, Issue 5, 105 (\textit{arXiv:} 1702.02800)

\bibitem{Mishra2018}
Mishra A and Chakraborty S 2018 \textit{Eur. Phys. J. C} \textbf{78}, Issue 5, 374 (\textit{arXiv:} 1710.06791)

\bibitem{Willenborg2018}
Willenborg F, Grunau S, Kleihaus B, and Kunz J 2018     \textit{Phys. Rev. D} \textbf{97} 124002 (\textit{arXiv:} 1801.09769)

\bibitem{Kratovitch2017}
Kratovitch P V, Potashov I M, Tchemarina Ju V, and Tsirulev A N 2017 \textit{Journal of Physics: Conference Series} \textbf{934}, Issue 1, 012047 (\textit{arXiv:} 1805.04447)

\bibitem{Nikonov2008}
Nikonov V V, Tchemarina Ju V and Tsirulev A N 2008 \textit{Class. Quantum Grav.} \textbf{25} 138001

\bibitem{BechmannLechtenfeld1995}
Bechmann O and Lechtenfeld O 1995
\textit{Class. Quantum Grav.} \textbf{12} 1473 (\textit{arXiv:} gr-qc/9502011)

\bibitem{BronnikovShikin2002}
Bronnikov K A and Shikin G N 2002 \textit{Grav. Cosmol.} \textbf{8} 107 (\textit{arXiv:} gr-qc/0109027)

\bibitem{Tchemarina2009}
Tchemarina Ju V and Tsirulev A N 2009
\textit{Grav. Cosmol.} \textbf{15} 94

\bibitem{Azreg-Ainou2010}
Azreg-A\"inou M 2010 \textit{Gen. Rel. Grav.} \textbf{42} 1427 (\textit{arXiv:} gr-qc/0912.1722)

\bibitem{Solovyev2012}
Solovyev D A and Tsirulev A N 2012
\textit{Class. Quantum Grav.} \textbf{29} 055013


\end{thebibliography}
\end{document}